\documentclass[useAMS,usenatbib]{mn2e}

\usepackage{graphicx}
\usepackage{caption}
\usepackage{amsmath}
\usepackage{amssymb}

\title[Unified line profiles for H perturbed by protons]{Unified line profiles for hydrogen perturbed by collisions with protons: satellites and asymmetries}
\author[Ingrid Pelisoli, M. G. Santos and  S. O. Kepler]{Ingrid Pelisoli$^{1}$\thanks{E-mail:
ingrid.pelisoli@ufrgs.br}, M. G. Santos$^{2}$ and S.O. Kepler$^{1}$ \\
$^{1}$Instituto de F\'{i}sica, Universidade Federal do Rio Grande do Sul, 91501-900 Porto Alegre, RS, Brazil\\
$^{2}$Campus Litoral Norte, Universidade Federal do Rio Grande do Sul, 90040-060 Porto Alegre, RS, Brazil}

\begin{document}

\date{Accepted 2015 January 22. Received 2015 January 21; in original form 2014 November 19}

\pagerange{\pageref{firstpage}--\pageref{lastpage}} \pubyear{2015}

\maketitle

\label{firstpage}

\begin{abstract}

We present new calculations of unified line profiles for hydrogen perturbed by collisions with protons. We report on new calculations of the potential energies and dipole moments which allow the evaluation of profiles for the lines of the Lyman series up to Lyman~$\delta$ and the Balmer series up to Balmer~10. Unified calculations only existed for the lines Lyman~$\alpha$ to Lyman~$\gamma$ and Balmer~$\alpha$ including the H$_2^+$ quasi-molecule. These data are available as online material accompanying this paper and should be included in atmosphere models, in place of the Stark effect of protons, since the quasi-molecular contributions cause not only satellites, but large asymmetries that are unaccounted for in models that assume Stark broadening of electrons and protons are equal.

\end{abstract}


\section{Introduction}

Collisional broadening occurs whenever a radiating particle interacts with a perturber during the process of emission or absorption of radiation. This interaction is known to cause not only a broadening of the line profile, but also the possibility of appearance of \textit{satellite-lines}. Satellites are explained by the formation, during the collision, of a transient bond state which has energy levels other than those of the non-interacting particle, producing a spectrum unlike the obtained for the isolated particle.

Hydrogen satellite-lines were already observed in some lines of the Lyman series in stars, and in the Lyman and the Balmer series in laboratory plasmas. In stars, these detections are mainly in spectra of white dwarf stars, but are also reported in A-type Horizontal Branch stars (which may be extremely low mass white dwarfs, ELMs) and $\lambda$ Bootis stars \citep{allard2000}. The first detections were two satellites in the red wing of Lyman~$\alpha$, close to $1400$ \citep{greenstein1980, wegner1982} and $1600$~\AA{} \citep{holm1985}, of DA-type white dwarfs with effective temperatures around $15\,000$~K. As they could not be explained by transitions of the hydrogen atom, which are the only lines expected to be observed in DA white dwarfs, they were explained by \citet{koester1985} and \citet{nelan1985} as \textit{quasi-molecular} contributions due to H$_2^+$ and H$_2$, respectively.

Since then, the effect of said two contributions has been included in atmosphere models, offering a good diagnostics of stellar parameters \citep[e.g.][]{allard1992, allard2004}. With the improvement of observational techniques and increase in data quality, satellites in other spectral lines have also been identified. \citet{koester1996} reported two strong absorptions in the red wing of Lyman~$\beta$, at $1060$ and $1078$~\AA, in the spectra of the DA white dwarf Wolf 1346 obtained with the \textit{Hopkins Ultraviolet Telescope} (HUT). In the same paper they already presented a model for the absorptions assuming a constant dipole moment for the transitions. \citet{allard1998} approached the problem again, now considering the true variation of the dipole moment. In Lyman~$\gamma$, \citet{wolff2001} identified a satellite with wavelength around $990$~\AA{} in the \textit{Far Ultraviolet Spectroscopic Explorer} (FUSE) spectrum of the star CD --38$^{\circ}$ 10980. Models were evaluated by \citet{allard2004b}, assuming constant dipole moments, and \citet{allard2009}, considering their dependence with distance.

In laboratory plasmas, experiments focused on quasi-molecular contributions were done for Lyman~$\alpha$ and the Balmer series. Lyman~$\alpha$ was studied by \citet{kielkopf1995} using laser-produced plasma in pure hydrogen. They detected five satellites: the already known features at $1400$ and $1600$~\AA{} and three others closer to the line centre, at $1234$, $1172$ and $1268$~\AA, explained by their models as contributions due to the quasi-molecules H$_2^+$ ($1234$~\AA) and H$_2$ ($1172$ and $1268$~\AA), but not detected in white dwarfs spectra because the large pressure broadening of the spectral lines hides them. \citet{kielkopf2002} studied Balmer~$\alpha$ and identified six satellites in the theoretical wing, of which only a stronger absorption at $8650$~\AA{} is observed in the plasma spectrum. For Balmer~$\beta$ and Balmer~$\gamma$, on the other hand, no satellites were identified, because only the line centre experimental profiles were published so far \citep{falcon2012, falcon2014} and usually there are no satellites in this region, which corresponds to radiation when the particles are far apart. In these papers and in \citet{falcon2010}, they use x-rays generated from a z-pinch dynamic hohlraum to drive plasma formation in a macroscopic gas cell and generate absorption spectra for the Balmer lines, with the intention of benchmarking white dwarf stars atmosphere models, though the published data are scarce to date.

Accuracy in model atmospheres for white dwarf stars is essential for a good determination of their physical parameters, especially effective temperature and surface gravity, as the fit of synthetic spectra is the most successful method to obtain those parameters \citep[e.g.][]{tremblay2009}. White dwarf stars are the final evolutionary state of all stars with masses below 8--10.5~M$_{\odot}$, depending on metallicity \citep[e.g.][]{iben1997, smartt2008, doherty2014}, corresponding to over 97~\% of the total number of stars. They generally do not present ongoing nuclear fusion and only emit their residual thermal energy, resulting in a slow cooling of the star. Because they have radii of the same order of the Earth's radius, they have a small surface area, so their cooling rate is small and therefore their cooling times are large; it takes approximately $10^{10}$ years for the effective temperature of a normal mass white dwarf to decrease from $100\,000$~K to near $5\,000$~K. As a result, the cool ones are among the oldest objects in the Galaxy, being extremely useful in the study of stellar formation and evolution history of the Milky Way \citep[e.g.][]{winget1987, bergeron1992, liebert2005, moehler2008, tremblay2014}.

Currently the atmosphere models in use to fit white dwarf stars and estimate their physical parameters only take into account quasi-molecular contributions for a few lines of the Lyman series, when they consider such contributions at all: Lyman~$\alpha$, Lyman~$\beta$ and Lyman~$\gamma$ \cite[e.g.][]{allard2009}. For the other lines of the Lyman series and the Balmer series, the interaction between hydrogen atom and free protons is approximated simply as electron Stark effect, which causes only a broadening that adds up to the effect of the electrons. The quasi-molecular effects are thus neglected, disregarding that they are responsible not only for a number of known satellites but also for strong asymmetries between the red and blue wings of all lines. The far red wing of Lyman~$\alpha$ ($\lambda>2\,500$~\AA), for example, turned out to be the missing opacity in the blue part of the optical spectrum of cool white dwarfs ($T_{\textrm{eff}}<6\,000$~K) of pure hydrogen composition. The inclusion of the broadening by collisions with H$_2$ was the key element in obtaining such result, as shown by \citet{kowalski2006}. However, contributions to other lines remain unaccounted for, despite the fact that their inclusion may be important to obtain accurate values for effective temperature and surface gravity.

This lack of accuracy results, for example, in a discrepancy in the calculated mean masses of DA white dwarfs when distinct methods are applied. \citet{falcon2010b}, analysing 449 DA white dwarfs from the \textit{European Southern Observatory Supernova Ia Progenitor Survey} (SPY), via the gravitational redshift method, obtained a mean mass of $\langle$M$_{\textrm{\tiny{DA}}}\rangle=0.647^{+0.013}_{-0.014}$~M$_{\odot}$. A consistent value of $\langle$M$_{\textrm{\tiny{DA}}}\rangle=0.630\pm0.028$~M$_{\odot}$ was obtained by \citet{romero2012} with seismology of 44 ZZ Ceti white dwarfs. On the other hand, when analysing $2\,217$ DA stars from the \textit{Sloan Digital Sky Survey} (SDSS) data release 7 with signal-to-noise ratio larger than 15 and $T_{\textrm{eff}}>13\,000$~K [to avoid the systematic super-estimative reported by \citet{kepler2006}], \citet{kleinman2013} found a mean mass of $\langle$M$_{\textrm{\tiny{DA}}}\rangle=0.593\pm0.002$~M$_{\odot}$, which is significantly smaller than the values obtained by \citet{falcon2010b} and \citet{romero2012}. Applying corrections to 3D convection, as calculated by \citet{tremblay2013}, \citet{kepler2015} obtained a mean mass of $0.662\pm0.003$~M$_{\odot}$ in the analysis of $1\,659$ DA white dwarfs from the SDSS data release 10 with signal-to-noise ratio larger than 15 and T$_{\textrm{eff}}>10\,000$~K, which is a value closer to the results of \citet{falcon2010b} and \citet{romero2012}. Hence the inconsistencies between the spectroscopic method and other methods are reduced when corrections to 3D convection are applied, but are still without a complete solution at present \citep{beaulieu2014}.
 
\citet{falcon2014} noted that the inferred electronic density of a plasma when fitting H~$\gamma$ was $\sim$ 40~\% smaller than the density obtained when fitting H~$\beta$, regardless of the set of models used. They suggested that this inconsistency would lead to lower determinations of surface gravities and, by consequence, a lower mean mass would be obtained. The probable cause for the different values obtained with each line is some missing physics in current atmosphere models. In this work we calculate line profiles for the quasi-molecular contributions of collisions between hydrogen and free protons, which is one of the effects not yet consistently included in the models. The potentials and dipole moments for H$_2^+$ used here are extensions of those published by \citet{santos2012}. Our results show that the quasi-molecular effects should be included in the atmospheric models, regardless of the detection or not of satellite lines, since they cause also large asymmetries in every line. As they are not contemplated in the current calculated spectra, they are probably affecting the determinations of temperature and gravity for the analysed data.

\section[]{Theoretical Basis}

There are basically four possible approaches for the calculation of collisional profiles: the impact theory, the quasi-static approximation (statistical theory), the unified theory and the quantum method. The impact theory \citep[e.g.][]{lindholm1945} assumes that the duration $\Delta t$ of a collision is much smaller than the interval between successive collisions ($\sim 2\pi/\Delta\omega$, where $\Delta\omega$ is the angular frequency measured from the line centre), being useful for determining the behaviour of the line centre, but not being able to account for the satellites. On the other extreme, there is the quasi-static approximation, which assumes $\Delta t \gg 2\pi/\Delta\omega$, being then a good method for obtaining the line wings \citep[e.g.][]{rohrman2011, santos2012}. \citet{anderson1952} showed that both approaches could be obtained from the same formulation, taken the appropriate limits, proposing the unified theory, which can give the shape of the whole line. \citet{allard1999} developed such theory in a quantum mechanical approach, obtaining from it better results for the shape and amplitude of the satellites. The pure quantum mechanical method can also give the whole line, but, for its implementation, one must know all of the states involved in a transition, including rovibrational, demanding larger computational time \citep{zygelman1990}. We chose the unified theory as proposed by \citet{allard1999} to the evaluations in this work.

In the unified theory, the normalized line profile measured from the line centre is given by the Fourier transform of an autocorrelation function $\Phi(s)$:
\begin{equation}
I(\Delta\omega) = \frac{1}{\pi} \textrm{Re} \int_0 ^{\infty} \Phi(s)\textrm{e}^{-\textrm{i}\Delta\omega s} \textrm{d}s \, .
\end{equation}
The autocorrelation function measures the influence of the presence of perturber particles -- in this case, free protons -- in the radiation of the particle -- here, hydrogen. As shown by \citet{allard1999}, it can be written as:
\begin{equation}
\Phi(s) = \textrm{e}^{ng(s)} \, ,
\end{equation}
where $n$ is the perturber density and $g(s)$, for a transition $\alpha=(i,f)$ between the initial state $i$ and the final one $f$, is given by
\begin{eqnarray}
g_{\alpha}(s) &=& \frac{1}{\sum_{e,e'}^{\alpha} |d_{ee'}|^2} \sum_{e,e'}{}^{\alpha}\int_{-\infty}^{\infty}\textrm{d}x_0\int_0^{\infty} 2\pi b \, \textrm{d}b \, \tilde{d}_{ee'}[R(0)] \nonumber \\
 && \times \left\{ \textrm{e}^{\frac{\textrm{i}}{\hbar}\int_0^s V_{ee'}[R(t)] \textrm{d}t} \tilde{d}^{\ast}_{ee'}[R(s)] - \tilde{d}_{ee'}[R(0)]\right\} \, , 
\end{eqnarray}
where $e$ and $e'$ denote the energy subspaces containing the states which approach, respectively, the initial and final atomic states of the transition as $R\rightarrow\infty$, with $R$ the distance between radiator and perturber, $d_{ee'}$ the dipole moments for the transition, $\tilde{d}_{ee'}$ the dipoles modulated by the the Boltzmann factor $\textrm{e}^{-\beta V_e}$, $x_0$ the perturber initial position, $b$ its impact parameter and $V_{ee'}(R)=V_{e}(R)-V_{e'}(R)$ the potential energy difference between states $i$ and $f$.

We adopted this approach to evaluate the line profiles $I(\Delta\omega)$ for the Lyman series up to Lyman~$\delta$ and to the Balmer series up to Balmer~$10$. The input data for potentials and dipole moments were calculated by \citet{santos2012} and are available at their web page\footnote{http://astro.if.ufrgs.br/marcios/index.htm}, except for the potential energy data used for the evaluation of the line centres, which were obtained with an improvement of the code, as will be described in Section \ref{center}. Calculations via unified theory existed in the literature only for Lyman~$\alpha$ \citep{allard2004}, Lyman~$\beta$ \citep{allard1998}, Lyman~$\gamma$ \citep{allard2009} and Balmer~$\alpha$ \citep{kielkopf2002}. For the other lines, calculations were done only by \citet{santos2012}, but via the quasi-static approach. Evaluations covering also the line centre were in order, not only to improve white dwarf atmosphere models but especially to allow comparison with new experimental data like those published by \citet{falcon2010, falcon2012, falcon2014}. That is the scope of the present work.

\section[]{Line Profiles for the Lyman Series}
\label{lyman}

\begin{figure}
\includegraphics[width=84mm]{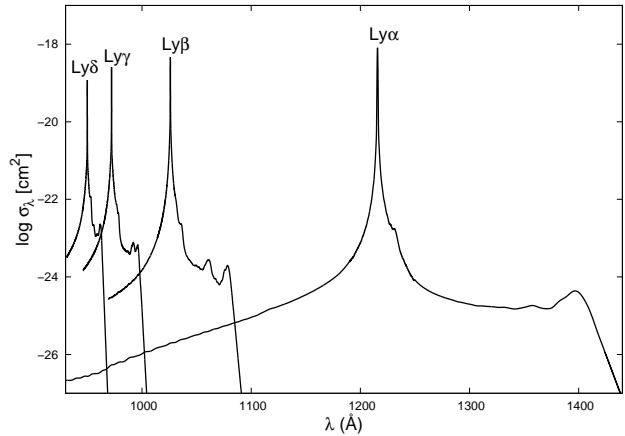}
\caption{Opacity profiles for every computed line in the Lyman series for a temperature of $10\,000$~K and perturber density of $10^{17}$~cm$^{-3}$. Strong satellites can be identified at the wavelengths of 1400, 1080, 1060, 990 and 965~\AA. The last one has not been identified in observations yet.}
\label{lyall}
\end{figure}

The Lyman series corresponds to transitions which have $n=1$ as the final level, for absorption lines, or as the initial level, in case of emission. Its wavelengths are in the ultraviolet region, between 912~\AA{} (Ly~$\infty$) and 1216~\AA{} (Ly~$\alpha$). Here we evaluated Ly~$\alpha$, Ly~$\beta$, Ly~$\gamma$ and Ly~$\delta$, whose opacity profiles are shown in Fig. \ref{lyall} for a temperature of $10\,000$~K and perturber density of $10^{17}$~cm$^{-3}$. The behaviour is linearly extrapolated for log($\sigma_{\lambda})\lesssim-26$. Many satellites can already be identified, as listed in Table \ref{lytab}, but each profiled is shown in more detail in the next subsections, for clarity.

\begin{table*}
 \centering
 \begin{minipage}{140mm}
  \caption{Wavelengths for the maximum of the stronger satellites in each calculated line in the Lyman series (for $T$~=~$10\,000$~K and $n$~=~$10^{17}$~cm$^{-3}$), the transition(s) that originates them and notes on observation and theoretical models.}
  \begin{tabular}{ccccc}
  \hline
  \\
  Line & Satellites (\AA) & Transition(s) & Observed? & Previous model\\
  \hline
             &         &                           &     & \\
  Ly$\alpha$ &  979.55 & 2p$\sigma_u$-3d$\pi_g$	   & no & \citet{allard1994}\footnote{\label{a1}via unified theory}\\
             & 1118.42 & 2s$\sigma_g$-2s$\sigma_g$ & no & \citet{allard1994}$^{\ref{a1}}$\\
             & 1230.88 & 1s$\sigma_g$-4f$\sigma_u$ & yes \citep{kielkopf1995} & \citet{kielkopf1995}$^{\ref{a1}}$\\
             &         & 1s$\sigma_g$-2p$\pi_u$    &     &                    \\
             & 1396.40 & 2p$\sigma_u$-3d$\sigma_g$ & yes \citep[e.g.]{koester1985} & e.g. \citet{allard1999}$^{\ref{a1}}$\\
             &         &                           &     & \\
  Ly$\beta$  & 1035.77 & 1s$\sigma_g$-6h$\sigma_u$ & no  & \citet{allard1998}$^{\ref{a1}}$\\
             & 1060.99 & 1s$\sigma_g$-4f$\pi_u$    & yes \citep{koester1996} & \citet{allard1998}$^{\ref{a1}}$\\
             & 1078.34 & 2p$\sigma_u$-5g$\pi_g$    & yes \citep{koester1996} & \citet{allard1998}$^{\ref{a1}}$\\
             &         &                           &     & \\
  Ly$\gamma$ &  978.24 & 1s$\sigma_g$-8k$\sigma_u$\footnote{The level 8k$\sigma_u$ is often miswritten in the literature as 8$j\sigma_u$. The letter $j$, by convention, must not used for orbital representation.} & no & \citet{allard2009}$^{\ref{a1}}$\\
             &  992.05 & 1s$\sigma_g$-8k$\sigma_u$  & yes \citep{wolff2001} & \citet{allard2009}$^{\ref{a1}}$\\
             &  996.03 & 2p$\sigma_u$-7i$\pi_g$     & yes \citep{wolff2001} & \citet{allard2009}$^{\ref{a1}}$\\
             &         &                            &    & \\
 Ly$\delta$  &  953.03 & 1s$\sigma_g$-10m$\sigma_u$ & no & \citet{santos2012}\footnote{\label{s1}via quasi-static approach} \\
             &         & 2p$\sigma_u$-9l$\pi_g$     &    & \\
             &  961.61 & 1s$\sigma_g$-8k$\pi_u$     & no & \citet{santos2012}$^{\ref{s1}}$ \\
             &         & 2p$\sigma_u$-9l$\sigma_g$  &    & \\
 \hline
\end{tabular}
\label{lytab}
\end{minipage}
\end{table*}

\subsection[]{Lyman~$\alpha$}

\begin{figure}
\includegraphics[width=84mm]{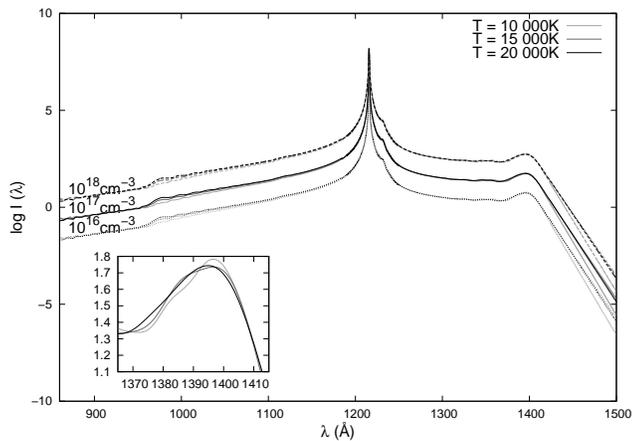}
\caption{Normalized profiles for the Lyman~$\alpha$ line. In black for $T$~=~$20\,000$~K, dark grey for $T$~=~$15\,000$~K and light grey for $T$~=~$10\,000$~K. The dotted line shows calculations for perturber density of $10^{16}$~cm$^{-3}$, the continuous for $10^{17}$~cm$^{-3}$ and the dashed for $10^{18}$~cm$^{-3}$. The same labels are adopted for the other Lyman lines. The smaller box shows in detail the behaviour of the satellite at $\sim$~$1400$~\AA{} with changes in temperature for density of $10^{17}$~cm$^{-3}$.}
\label{ly2}
\end{figure}

Lyman~$\alpha$ has contributions due to six transitions of the H$_2^+$ quasi-molecule. The total profile accounting for each contribution is shown in Fig. \ref{ly2} for temperatures of $10\,000$, $15\,000$ and $20\,000$~K and perturber densities of $10^{16}$, $10^{17}$ and $10^{18}$~cm$^{-3}$. Four satellites can be identified, with approximate wavelengths of 980, 1100, 1240 and 1400~\AA. The first two are probably too faint to be detected in stellar spectra, at least at these temperatures. The one at 980~\AA{} would also be blended with the Lyman~$\gamma$ line in stellar spectra. The other two were already observed, as cited in the introduction, in laboratory plasmas and in white dwarfs. We can also note that, in large scale, the temperature does not seem to affect the profiles, but, in the detailed box, it can be seen that the satellite at $\sim$~$1400$~\AA{} has a consistent behaviour: it broadens with the increase in temperature, while its peak is displaced to smaller wavelengths. Such behaviour might be useful as a good temperature diagnostic for plasmas. The effect of increasing density is simply a linear shift (in log scale) and a larger broadening.

\begin{figure}
\includegraphics[width=84mm]{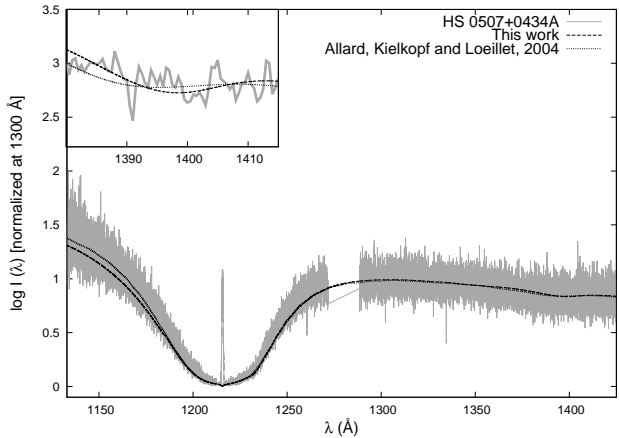}
\caption{HST spectrum of the white dwarf HS~0507+0434A (grey) and \textsc{Tlusty}/\textsc{Synspec} models using the quasi-molecular contribution tables evaluated by us (dashed line) and by N. F. Allard (dotted line), assuming temperature of $20\,000$~K and log(g)~=~7.75 [close to the values determined by \citet{andrews2012}]. In the small box, the spectrum was smoothed to enhance the position of the satellite.}
\label{hs0507}
\end{figure}

\begin{figure}
\includegraphics[width=84mm]{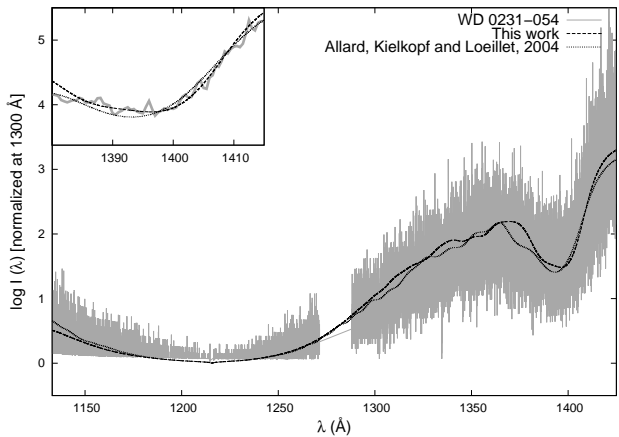}
\caption{HST spectrum of the white dwarf WD~0231-054 (grey) and \textsc{Tlusty}/\textsc{Synspec} models using the quasi-molecular contribution tables evaluated by us (dashed line) and by N. F. Allard (dotted line), assuming temperature of $15\,000$~K and log(g)~=~8.80 [similar to the values determined by \citet{gianninas2011}]. Again we show the smoothed spectrum in the satellite region to enhance the satellite position, which is better fitted by our data.}
\label{wd0231-054}
\end{figure}

In Figs. \ref{hs0507} and \ref{wd0231-054}, we show comparison between observed data for the stars HS~0507+0434A and WD~0231-054 in the region of Lyman~$\alpha$ and calculated models using the codes \textsc{Tlusty} \citep{hubeny1988} and \textsc{Synspec} \citep{hubeny2011}. The dashed line shows the model obtained using the quasi-molecular contributions calculated in this work, and the dotted line uses tables provided by N. F. Allard\footnote{http://mygepi.obspm.fr/$\sim$allard/}. The models are not fitted to the data, but simply over-plotted considering normalization at an arbitrary wavelength and assuming effective temperature and surface gravity found in the literature. In both cases, there are two major differences between the models: the blue wing of the central line and the position of the satellite. Allard's models seem to fit better the central line, but the central wavelength of the satellite is underestimated, being better determined by our model. A possible explanation for such difference is that Allard's calculations use potential energy data only to small distances, so the real asymptotic value of each level is not reached. As a subtraction of the calculated asymptotic value [so that V(R$_{\textrm{max}}$)~=~0] is necessary to the numerical evaluation of the line, a larger value than the real one is subtracted, overestimating the shift in the difference potential and making the satellite appear closer to the line centre. Such problem does not happen in our calculations, since we evaluated the potential energy data to large distances.

\subsection[]{Lyman~$\beta$}

\begin{figure}
\includegraphics[width=84mm]{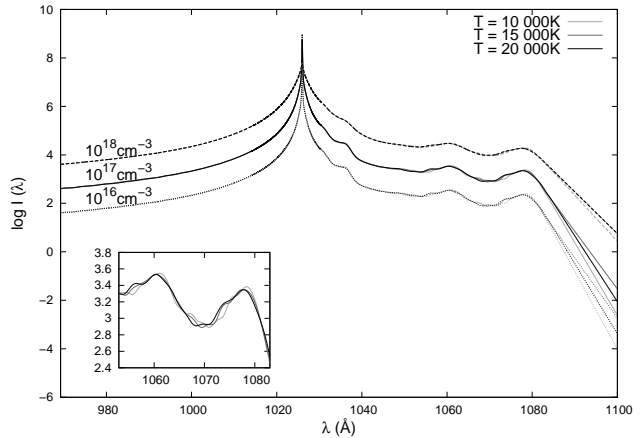}
\caption{Normalized profiles for the Lyman~$\beta$ line. The smaller box shows in detail the behaviour of the two stronger satellites, at 1060 and 1078~\AA, with changes in temperature for density $10^{17}$~cm$^{-3}$.}
\label{ly3}
\end{figure}

Ten transitions contribute to the Lyman~$\beta$ profile. Their combined effect gives the total profile shown in Fig. \ref{ly3}. Our calculations are shown for temperatures of $10\,000$, $15\,000$ and $20\,000$~K and perturber densities of $10^{16}$, $10^{17}$ and $10^{18}$~cm$^{-3}$. Apart from the strong satellites at 1060 and 1078~\AA{} already studied in white dwarf spectra, there is another satellite around 1035~\AA, probably too close to the line centre to be identified in white dwarfs, since the broadening of the line due to other mechanisms hides it. Temperature is again not an important factor, result that is current in the literature. However, again in a more detailed scale, the strong satellites do show some variation. They also show small oscillations. Possible reasons for that are the fact that the calculations presented here assume that the perturbers have a single constant velocity and no averaging was done, causing an effect similar to diffraction, as mentioned by \citet{allard1978} and, mainly, the contribution of the window function in the Fourier transform [as reported by \citet{allard1974}].

\subsection[]{Lyman~$\gamma$}

\begin{figure}
\includegraphics[width=84mm]{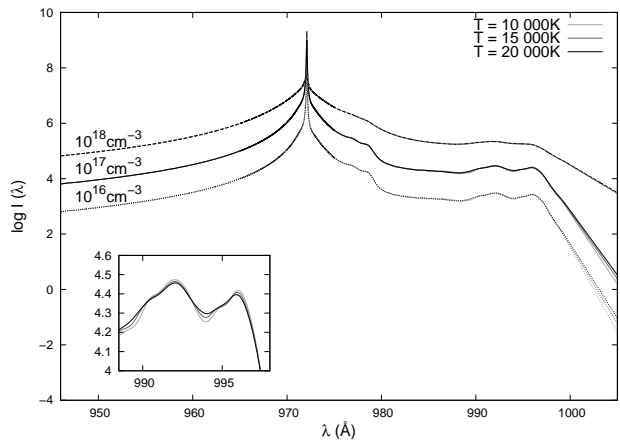}
\caption{Normalized profiles for the Lyman~$\gamma$ line. The smaller box shows in detail the behaviour of the satellites at 990 and 995~\AA{} with changes in temperature for density $10^{17}$~cm$^{-3}$.}
\label{ly4}
\end{figure}

Fourteen transitions of the H$_2^+$ quasi-molecule contribute to Lyman~$\gamma$. The calculated total profiles, for temperatures of $10\,000$, $15\,000$ and $20\,000$~K and perturber densities of $10^{16}$, $10^{17}$ and $10^{18}$~cm$^{-3}$, are shown in Fig. \ref{ly4}. The noticeable satellite features are at 979, 990 and 995~\AA. The first one is again to close to the line centre for detection in stars, but the combined effect of the other two was already detected, as detailed in the introduction. Since this satellite has two components, as already reported by \citet{allard2009}, its behaviour with temperature shall not be very simple, because each component has its independent variation. The rest of the line shows no significant change.

\begin{figure}
\includegraphics[width=84mm]{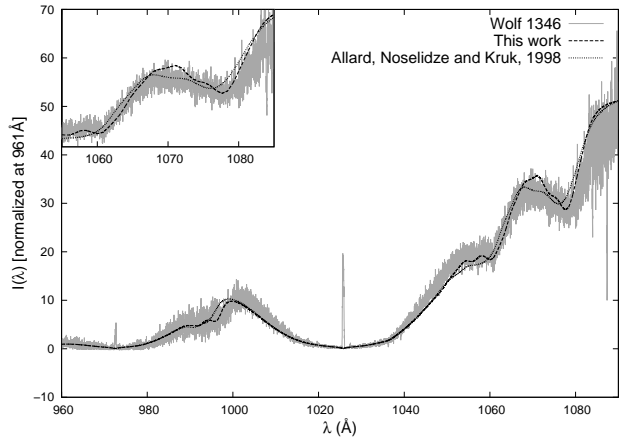}
\caption{FUSE spectrum of the white dwarf Wolf~1346 (grey) and \textsc{Tlusty}/\textsc{Synspec} models using the quasi-molecular contribution tables evaluated by us (dashed line) and by N. F. Allard (dotted line), assuming temperature of $20\,700$~K and log(g)~=~8.00 [as determined by \citet{giammichele2012}].}
\label{wolf1346}
\end{figure}

\begin{figure}
\includegraphics[width=84mm]{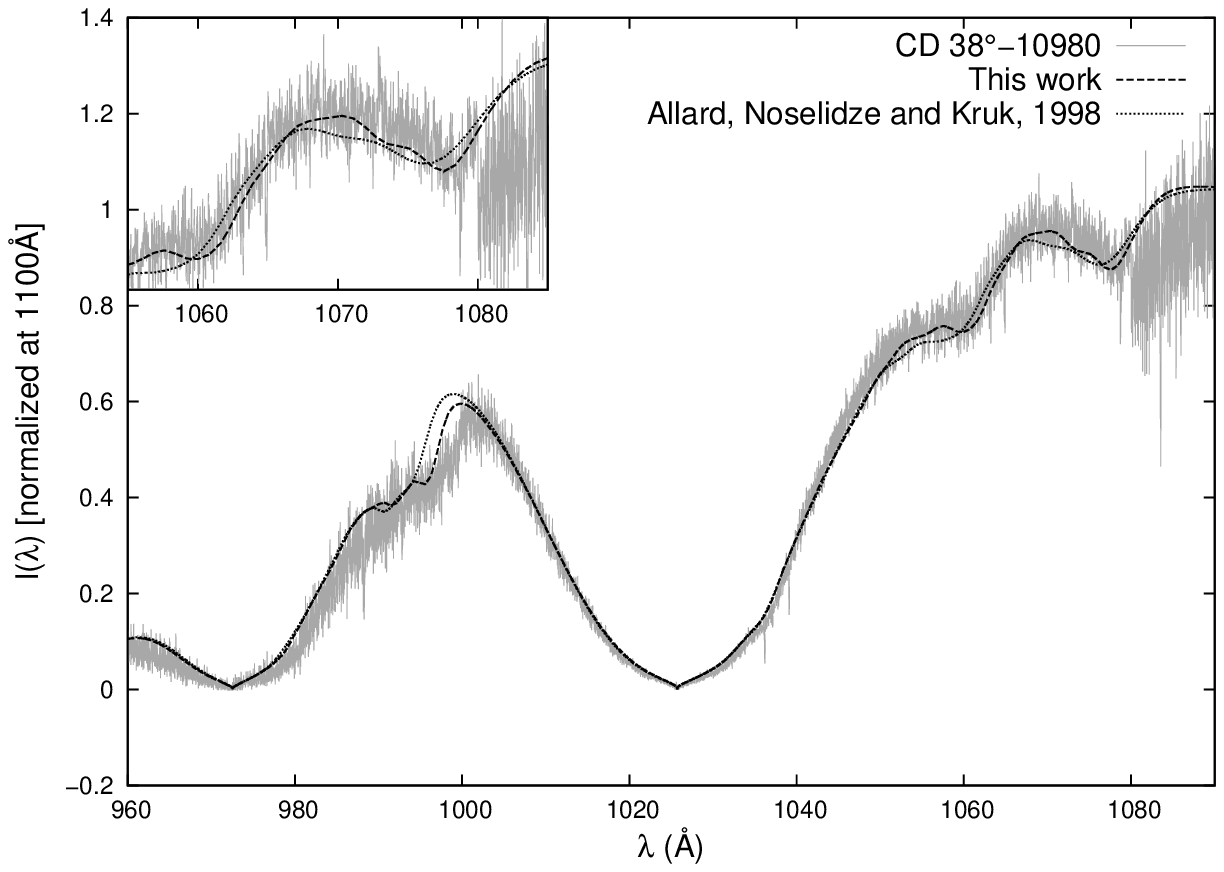}
\caption{FUSE spectrum of the white dwarf CD-38$^{\circ}$10980 (grey) and \textsc{Tlusty}/\textsc{Synspec} models using the quasi-molecular contribution tables evaluated by us (dashed line) and by N. F. Allard (dotted line), assuming temperature of $26\,000$~K and log(g)~=~8.00 [as determined by \citet{giammichele2012}].}
\label{cd3810980}
\end{figure}

In Figs. \ref{wolf1346} and \ref{cd3810980}, we show FUSE observed spectra in the region of the lines Lyman~$\beta$ and Lyman~$\gamma$ compared to calculated models using the codes \textsc{Tlusty} \citep{hubeny1988} and \textsc{Synspec} \citep{hubeny2011}. As before, the dashed line shows the model which uses the quasi-molecular contributions calculated in this work and the dotted line uses tables by N. F. Allard. In this case, there seems to be no difference between the central line of each model, but the position of the satellite is again underestimated by Allard's models, probably due to the same reason as for the Lyman~$\alpha$ satellite. The underestimation becomes more evident as one moves to high-order lines because higher levels converge at larger distances, so that the higher the line, the larger is the error in the calculated asymptotic value and, by consequence, the larger the shift of the satellite towards the central line.

\subsection[]{Lyman~$\delta$}

\begin{figure}
\includegraphics[width=84mm]{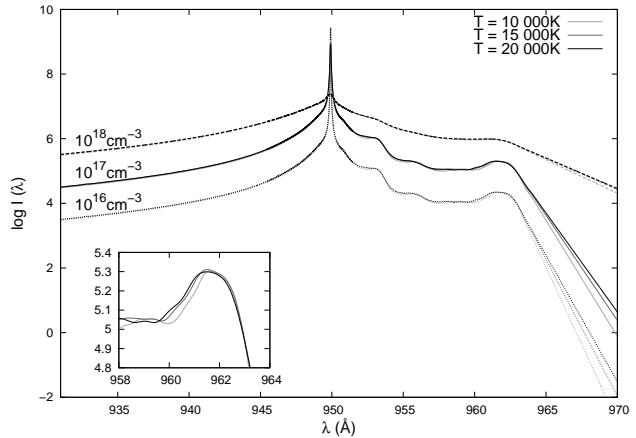}
\caption{Normalized profiles for the Lyman~$\delta$ line. The smaller box shows in detail the behaviour of the satellite at 962~\AA{} with changes in temperature for density $10^{17}$~cm$^{-3}$.}
\label{ly5}
\end{figure}

Lyman~$\delta$ has contributions of 18 transitions. It is evaluated with the unified theory for the first time in this paper. The normalized profiles are shown in Fig. \ref{ly5} for temperatures of $10\,000$, $15\,000$ and $20\,000$~K and perturber densities of $10^{16}$, $10^{17}$ and $10^{18}$~cm$^{-3}$. We find stronger satellites around 953 and 962~\AA. Neither has reported detection in the literature. The former probably is to close to the line centre for that, but the latter, although not mentioned in the literature, is present in the FUSE spectrum of CD --38$^{\circ}$10980 published in fig. 4 of \citet{wolff2001}. Similar to what was noted for the $1400$~\AA{} satellite in Ly~$\alpha$, we see that its peak moves to smaller densities and that it broadens as temperature increases, so its behaviour is probably also a good diagnostic for temperature determinations.

\section[]{Line Profiles for the Balmer Series}
\label{balmer}

\begin{figure}
\includegraphics[width=84mm]{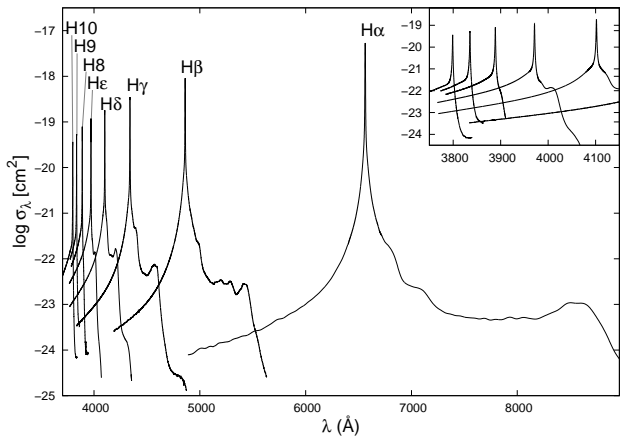}
\caption{Opacity profiles for every computed line in the Balmer series for a temperature of $10\,000$~K and perturber density of $10^{17}$~cm$^{-3}$. The satellites are not very intense in this case, but it is worth noticing that all lines are very asymmetrical, especially the high order ones (in the detailed box), in which the red wing falls of very quickly, while the blue wing has a smother behaviour.}
\label{baall}
\end{figure}

The Balmer series is composed by transitions that have $n=2$ as the final level, in case of absorption, or the initial level, for emission. It is located in the visible portion of the spectrum, between 3646~\AA{} (H~$\infty$) and 6563~\AA{} (H~$\alpha$). We evaluate here every line from H~$\alpha$ to H~$10$, as can be seen in the opacity profiles in Fig. \ref{baall}, for temperature of $10\,000$~K and perturber density of $10^{17}$~cm$^{-3}$. The stronger satellites are listed in Table \ref{batab}. Only one of them was observed so far, which is probably due to lack of laboratory data away from the line centre for the other lines. Such satellites would be stronger for temperatures around $10\,000$~K, but the ionization rate of hydrogen is low at such temperature, so they are effectively stronger around $15\,000$~K, when the H$^{+}$ population is more significant. In white dwarfs, lines above H~$11$ are hardly ever seen, since the high pressure causes then to overlap and disappear. The profiles evaluated here are probably sufficient to study any spectra in the visible region.

\begin{table*}
 \centering
 \begin{minipage}{140mm}
  \caption{Wavelengths for the maximum of the stronger satellites (if existent) in each calculated line in the Balmer series (for $T$~=~$10\,000$~K and $n$~=~$10^{17}$~cm$^{-3}$), the transition(s) that originates them and notes on observation and previous theoretical models.}
  \begin{tabular}{ccccc}
  \hline
  \\
  Line & Satellite (\AA) & Transition(s) & Observed? & Previous model\\
  \hline
             &         &                        &     & \\
  H$\alpha$  & 8506.60 & 3d$\pi_g$-4f$\pi_u$    & yes \citep{kielkopf2002} & \citet{kielkopf2002}\footnote{via unified theory} \\
             &         & 4f$\sigma_u$-5g$\pi_g$ &     & \\
             &         &                        &     & \\
  H$\beta$   & 5413.33 & 3d$\pi_g$-6h$\pi_u$    & no & \citet{santos2012}\footnote{\label{s2}via quasi-static approach} \\
             &         &                        &    & \\
  H$\gamma$  & 4392.21 & 3d$\pi_g$-8k$\delta_u$ & no & \citet{santos2012}$^{\ref{s2}}$ \\
             &         & 3d$\pi_g$-9k$\sigma_u$ &    & \\
             &         & 3p$\sigma_u$-7g$\sigma_g$ & & \\
             & 4571.88 & 2p$\sigma_u$-7i$\delta_g$ & no & \citet{santos2012}$^{\ref{s2}}$ \\
             &         & 3d$\sigma_g$-8k$\pi_u$    &   & \\
             &         & 3d$\pi_g$-8k$\pi_u$    &   & \\
             &         &                        &    & \\
 H$\delta$   & 4203.87 & 2p$\sigma_u$-10l$\sigma_g$ & no & \citet{santos2012}$^{\ref{s2}}$ \\
             &         & 3d$\pi_g$-9k$\pi_u$        &  & \\
             &         &                        &    & \\
 H$\epsilon$ & 4005.35 & 2s$\sigma_g$-10k$\pi_u$ & no & \citet{santos2012}$^{\ref{s2}}$ \\
             &         & 2p$\sigma_u$-10i$\delta_g$ & & \\
             &         & 2p$\sigma_u$-11l$\delta_g$ & & \\
             &         & 2p$\sigma_u$-12n$\sigma_g$ & & \\
             &         & 3p$\sigma_u$-12n$\sigma_g$ & & \\
             &         & 3d$\pi_g$-10k$\pi_u$ & & \\
             &         & 4f$\sigma_u$-12n$\sigma_g$ & & \\
\hline
\end{tabular}
\label{batab}
\end{minipage}
\end{table*}

\subsection[]{Balmer~$\alpha$}

\begin{figure}
\includegraphics[width=84mm]{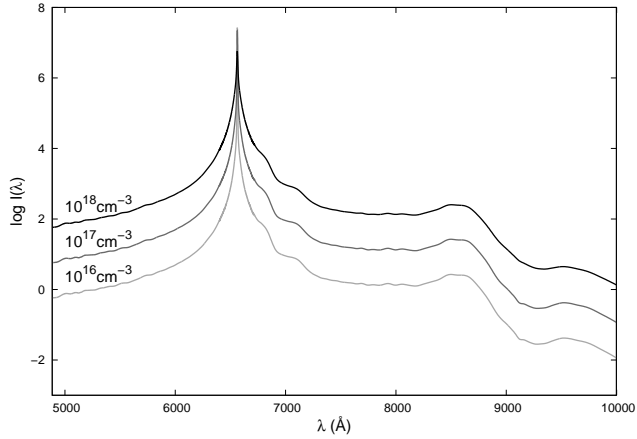}
\caption{Normalized profiles for the Balmer~$\alpha$ line. The temperature is $10\,000$~K and densities are $10^{16}$ (\textit{light grey}), $10^{17}$ (\textit{dark grey}) and $10^{18}$~cm$^{-3}$ (\textit{black}). The same labels are adopted for the other Balmer lines. Various satellites appear in the red wing, making the line very asymmetric.}
\label{ba3}
\end{figure}

For the Balmer~$\alpha$ line, there are contributions due to 32 transitions. The total profile is shown in Fig. \ref{ba3} for a temperature of $10\,000$~K and densities of $10^{16}$, $10^{17}$ and $10^{18}$~cm$^{-3}$. The strongest satellite is around 8500~\AA{} and has been observed in plasma spectra by \citet{kielkopf1995}. However, in white dwarfs, it has not been identified for two reasons: first, it is too close to the region were the Paschen series is and, second, because, for temperatures at which the H$_2^+$ population is relevant ($10\,000$--$25\,000$~K), the star emits predominantly in the blue, having little flux in the region of this satellite. The satellite also contributes to the line asymmetry.

\subsection[]{Balmer~$\beta$}

\begin{figure}
\includegraphics[width=84mm]{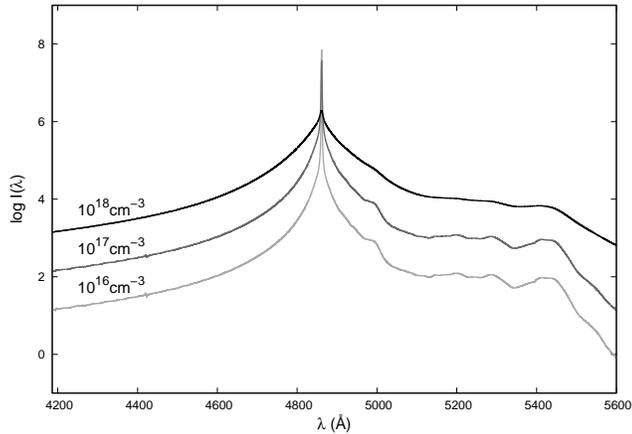}
\caption{Normalized profiles for the Balmer~$\beta$ line. The stronger satellite is at 5400~\AA, but it is still too faint for observation, its main importance being thus the asymmetry it causes.}
\label{ba4}
\end{figure}

Forty-six transitions contribute to Balmer~$\beta$. Its total profile is shown in Fig. \ref{ba4} for a temperature of $10\,000$~K and densities of $10^{16}$, $10^{17}$ and $10^{18}$~cm$^{-3}$. We can see many faint satellites in the red wing, causing asymmetry. The strongest satellite is around 5400~\AA, but it is probably too faint compared to the line intensity to be observed in white dwarfs. It is important to notice, though, that it contributes to the asymmetry, which, is worth remembering, does not appear if the interaction between hydrogen and proton is treated simply as Stark effect, as is done in current atmosphere models.

\subsection[]{Balmer~$\gamma$}

\begin{figure}
\includegraphics[width=84mm]{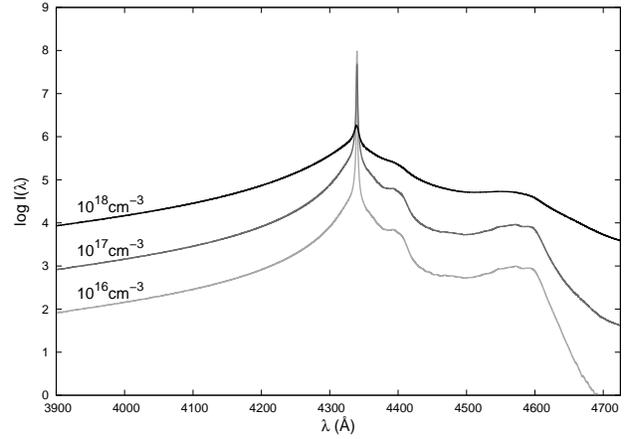}
\caption{Normalized profiles for the Balmer~$\gamma$ line. There are two strong satellite features, at 4400 and 4600~\AA, none of them reported in observed spectra until now.}
\label{ba5}
\end{figure}

To Balmer~$\gamma$ there are contributions due to 60 transitions. The total profile is shown in Fig. \ref{ba5} for a temperature of $10\,000$~K and perturber densities of $10^{16}$, $10^{17}$ and $10^{18}$~cm$^{-3}$. There are two noticeable satellites, close to 4400 and 4600~\AA. The former is too close to the line centre to be observed in white dwarfs, the latter, despite being away from the centre, has very small intensity over the continuum, what would difficult or even prevent its observation. However, both of them cause asymmetry.

\subsection[]{Balmer~$\delta$}

\begin{figure}
\includegraphics[width=84mm]{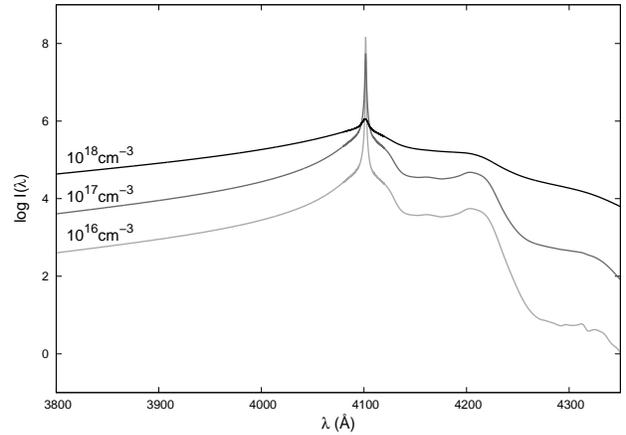}
\caption{Normalized profiles for the Balmer~$\delta$ line. The most noticeable satellite is at 4200~\AA, but again it seems too faint to be observed in white dwarf spectra.}
\label{ba6}
\end{figure}

We have to account for contributions of 74 transitions in Balmer~$\delta$. Its profile for a temperature of $10\,000$~K and perturber densities of $10^{16}$, $10^{17}$ and $10^{18}$~cm$^{-3}$ can be seen in Fig. \ref{ba6}. The strongest satellite is close to 4200~\AA, but, as for the 4600~\AA{} satellite in Balmer~$\gamma$, its intensity does not seem to be strong enough, when compared to the continuum, to allow its detection in white dwarf spectra.

\subsection[]{Balmer~$\epsilon$}

\begin{figure}
\includegraphics[width=84mm]{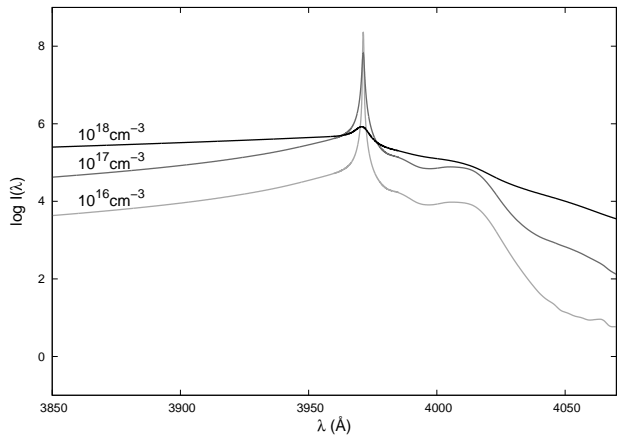}
\caption{Normalized profiles for the Balmer~$\epsilon$ line. A satellite feature appears at 4005~\AA, and the intensity of the profile drops after it.}
\label{ba7}
\end{figure}

There are 88 transitions that contribute to Balmer~$\epsilon$. We show the total profile in Fig. \ref{ba7} for a temperature of $10\,000$~K and perturber densities of $10^{16}$, $10^{17}$ and $10^{18}$~cm$^{-3}$. We can see a very faint satellite around 4005~\AA, with intensity of the same order of the continuum in the blue wing of the line, meaning it is probably not observable. However, as the intensity of the red wing rapidly decreases after the satellite, while the intensity in the blue line changes slowly, knowing the satellite is important to model the line asymmetry.

\subsection[]{Balmer~$8$, Balmer~$9$ and Balmer~$10$}

\begin{figure}
\includegraphics[width=84mm]{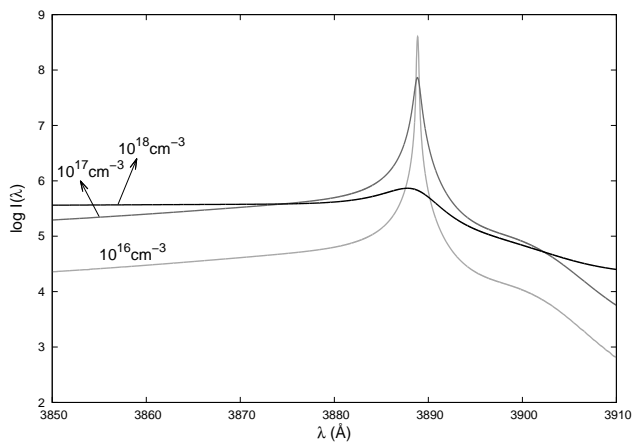}
\caption{Normalized profiles for the Balmer~$8$ line. No satellites are seem, but the line is largely asymmetric.}
\label{ba8}
\end{figure}

\begin{figure}
\includegraphics[width=84mm]{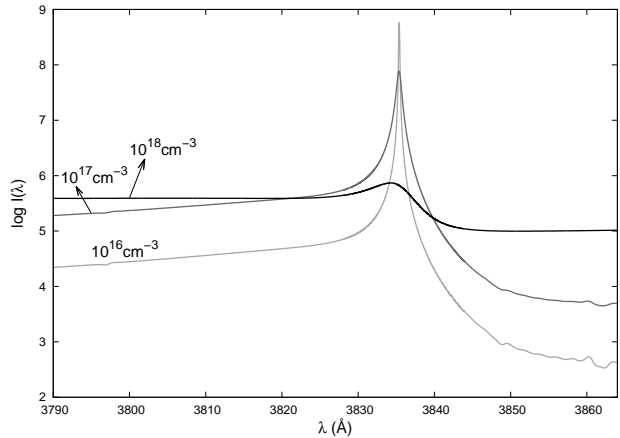}
\caption{Normalized profiles for the Balmer~$9$ line. In spite of the absence of satellites, we can note a line asymmetry that does not exist in pure-Stark profiles.}
\label{ba9}
\end{figure}

\begin{figure}
\includegraphics[width=84mm]{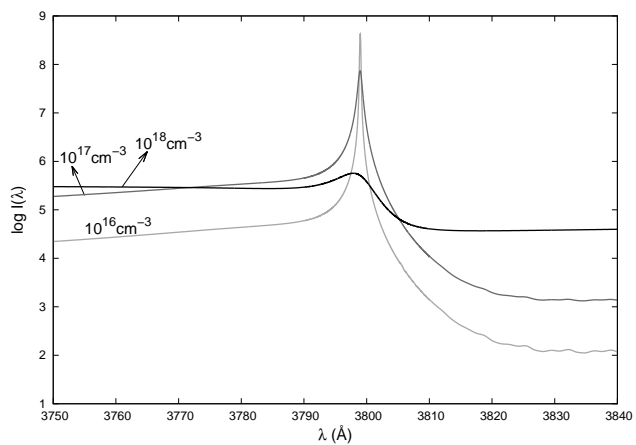}
\caption{Normalized profiles for the Balmer~$10$ line. Again no satellites, but a large asymmetry between red and blue wings can be seen.}
\label{ba10}
\end{figure}

The lines of Balmer~$8$, Balmer~$9$ and Balmer~$10$ have contributions due to 102, 116 and 130 transitions, respectively. Their profiles are shown in Fig. \ref{ba8}, Fig. \ref{ba9} and Fig. \ref{ba10}, again for a temperature of $10\,000$~K and perturber densities of $10^{16}$, $10^{17}$ and $10^{18}$~cm$^{-3}$. It is worth noticing that, for $n$~=~$10^{18}$cm$^{-3}$, these high order lines start to fade due to the high pressure. There are no satellites visible in those lines. One might wonder in that case if the quasi-molecular contributions really matter. However, it is very clear that they \textit{do} affect the line behaviour, again causing a strong asymmetry between the red and blue wings, which is until now completely ignored in white dwarf stars atmosphere models.

\section[]{Comparison to Electronic Stark}

In current synthetic spectra for DA white dwarf stars, the effect of the interaction between free protons and radiating hydrogen is assumed to be exactly the same of the interaction of free electrons with hydrogen. So the profiles are calculated with density $n_e$ of electrons, and such density is simply doubled to account for protons as well, as the number of free protons must be the same as the number of free electrons, because both originate, in this case, when hydrogen is ionized. As already said in the introduction, such approach disregards the quasi-molecular contributions that appear when temperature and pressure are adequate to the formation of a transient bond state between hydrogen and proton.

\begin{figure}
\includegraphics[width=84mm]{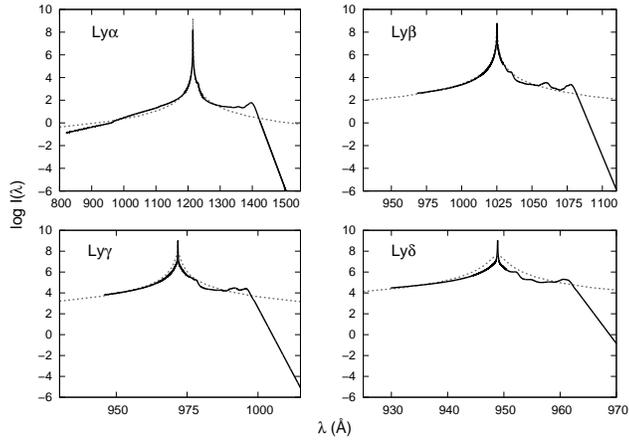}
\caption{Comparison between each calculated line in the Lyman series (\textit{black continuous line}) and the correspondent pure electronic Stark profile (\textit{grey dashed line}) for $T$~=~$10\,000$~K and log($n_e$)~=~log($n_p$)~=~17. There is no great difference if one considers only the blue wing, but the red wing is very different, showing satellite features when the quasi-molecular contributions are taken into account.}
\label{stark-ly}
\end{figure}

\begin{figure}
\includegraphics[width=84mm]{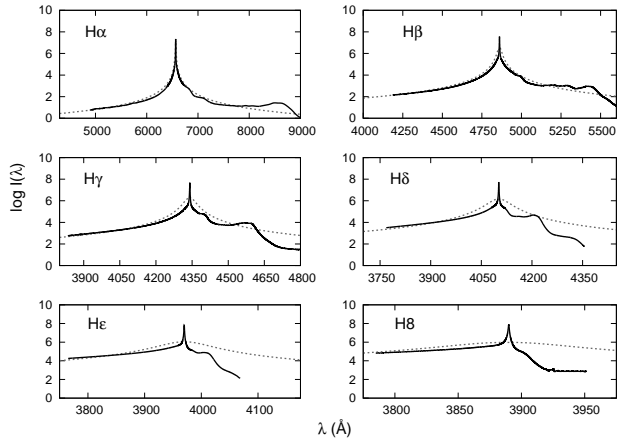}
\caption{Comparison between calculated lines in the Balmer series (\textit{black continuous line}) and the correspondent pure electronic Stark profile (\textit{grey dashed line}) for $T$~=~$10\,000$~K and log($n_e$)~=~log($n_p$)~=~17. For the lower order lines, we see that the satellite in the blue wing must have a considerable contribution. On the other hand, in the high order lines, the major difference is the rapid decreasing of the red wing compared to the blue wing, effect that does not appear in the electronic Stark profiles.}
\label{stark-ba}
\end{figure}

This is not a good practice and must be affecting the determination of temperature and surface gravity. We plot here the calculated lines for $T$~=~$10\,000$~K and log($n$)~=~17 together with the correspondent electronic Stark effect to show that. Profiles for the Lyman series are shown in Fig. \ref{stark-ly} and for the Balmer series in Fig. \ref{stark-ba}. It is easy to notice two large differences: the existence of satellites and the large asymmetries between red and blue wings. The former is more important for low order lines (n$_f \leq$ 5), when the red wing usually shows at least one strong satellite. The latter is evident for high order lines (n$_f >$ 5), in which the red wing falls quickly for the quasi-molecular profile, while the red wing in the electronic Stark behaves similarly to the blue wing and decreases slowly. These differences may be the cause of the inconsistent densities reported by \citet{falcon2014} and \citet{beaulieu2014}, which are obtained when different lines are fitted.

It is worth mentioning that the validity domain of the VCS theory, in which the Stark models shown here are based, is $\Delta\omega/\omega_p \leq 1$ \citep{vcs1971}, where $\Delta\omega$ is the angular frequency measured from the line centre and $\omega_p=\sqrt{4\pi n_e e^2/m_e}$ is the plasma frequency (\textit{cgs} units). According to that, the profiles would be valid only in the central 1--10~\AA{} of the line. However, the validity of the theory has been extrapolated in the current models, since observed lines of white dwarf stars are usually more than 100~\AA{} wide.

\section{The line centre}
\label{center}

Quasi-static profiles are inherently a poor approximation to the line centre, since its main assumption of zero velocity implies that only collisions whose duration is larger than the interval between them are taken into account, and that is only appropriate for the frequencies in the line wings. Therefore, the central behaviour of the profiles evaluated by \citet{santos2012} should not be relied upon. Besides that, such profiles were calculated summing up contributions due to different transitions, when, actually, they should be convoluted, since lifetimes for hydrogen excited states are of the order of $10^{-8}$~s, while the typical duration of collisions in the physical conditions of white dwarfs is around $10^{-13}$--$10^{-12}$~s, implying that different states may form throughout the electronic transition and contribute to the line.

The profiles calculated here are based on the unified theory, in which no assumption is made concerning the duration of collisions, and the partial profiles are convoluted, so the form of the line can be relied upon for the whole wavelength range. Nevertheless, we recall that both the unified theory and the quasi-static approach assume adiabatic perturbations, disregarding interference between different levels, which can result in underestimation of the line shift \citep{allard1999}. However, as the behaviour of the line centre in white dwarf stars is dominated by electronic Stark and Doppler effects, being the quasi-molecular contributions more important in the wings, the profiles shown in the previous sessions should be satisfactory for model atmosphere applications. 

\begin{figure}
\includegraphics[width=84mm]{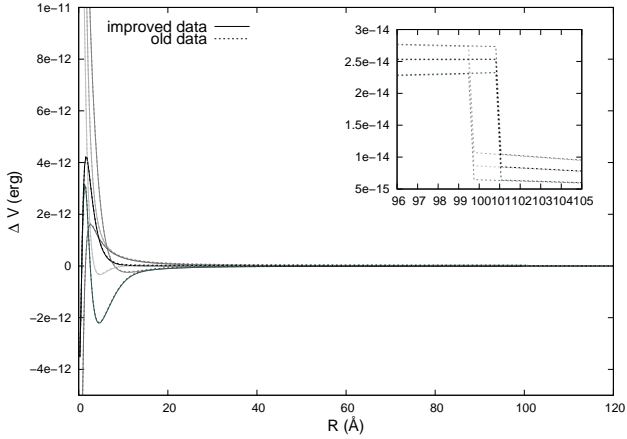}
\caption{Difference potentials (minus the asymptotic value) for the six transitions contributing to the Lyman-$\alpha$ line. We can note that in the old data there is a discontinuity, which, transformed to wavelength, corresponds to $>$1.2~\AA. This issue was solved in the new data.}
\label{potentials}
\end{figure}

\begin{figure}
\includegraphics[width=84mm]{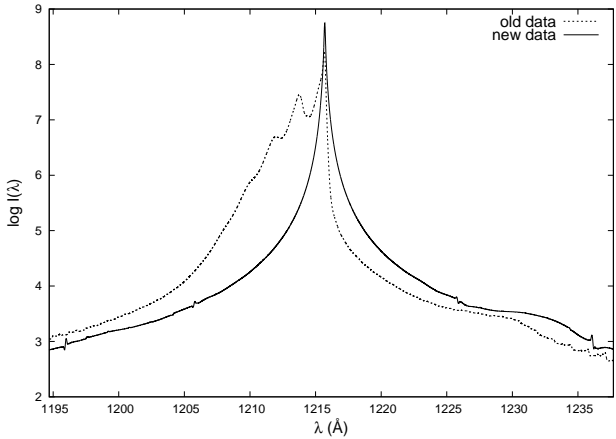}
\caption{Profile for the centre of the Lyman~$\alpha$ line for a temperature of $10\,000$~K and perturber density of 10$^{17}$~cm$^{-3}$ evaluated using the data published in \citet{santos2012} (\textit{dashed line}) and the profile calculated with the improved data (\textit{continuous line}). We can note an abnormal behaviour in the old profiles, which is a consequence of discontinuities in the potential data.}
\label{center1}
\end{figure}

\begin{figure}
\includegraphics[width=84mm]{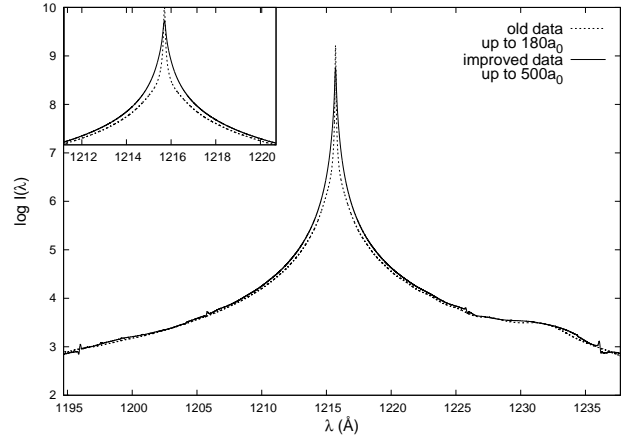}
\caption{Same as Fig. \ref{center1}, but now truncating the old potentials before the discontinuity. The abnormal behaviour disappears, but as the asymptotic behaviour of the potential is not reached, the contributions to the centre are not so well determined. A better determination is obtained with the improved data.}
\label{center2}
\end{figure}

However, with the advent of experiments such as the one done by \citet{falcon2012} at the Z Pulsed Power Facility at Sandia National Laboratories, more accurate profiles in the central region of the line are necessary. When trying to evaluate such detailed profiles using the published data of \citet{santos2012}, we noticed that the data was not accurate enough for that. There were discontinuities in some of the potential data, due to convergence issues in the implemented computational methods, as can be seen on Fig. \ref{potentials}. That lead to abnormal behaviour in the line centre, as shown for Lyman~$\alpha$ in Fig. \ref{center1}. To avoid that behaviour, it was necessary to truncate the potentials at $\sim$180~$a_0$, where a$_0\approx0.5292$~\AA{} is the Bohr radius, but that way the asymptotic behaviour is not reached and the line centre is still not accurately determined (Fig. \ref{center2}), since it is formed mainly when the particles are far apart. Besides, as the mean distance between the particles for density of 10$^{17}$cm$^{-3}$, for example, is around 250~a$_0$, going up to 180~a$_0$ is not sufficient if one is concerned with the line centre. As at that distance the asymptotic behaviour of the potential energy, which is $\propto r^{-2}$, is still not reached, it is also not possible to extrapolate the data with the necessary precision. For larger values of $n$, the asymptotic behaviour is reached even farther away, so the truncation may largely affect the determination of the line centre profile.

Thus, we recalculated the potential data for the H$_2^+$ quasi-molecule. In order to check the convergence of the potentials to their asymptotic values, we had to make use of a larger number of terms in the recurrence relations\footnote{In \citet{santos2012}, the recurrence relations were truncated at indexes equal to 25. Now, for n=1 and n=2 for example, we had to truncate the indexes at 300 to get the correct asymptotic value to a precision at the 21$^{\textrm{st}}$ decimal.}, what leads to a larger set of homogeneous linear equations and demands, in consequence, larger matrices to solve. We calculated the profiles for Lyman~$\alpha$, Lyman~$\beta$ and Balmer~$\alpha$ with the new data.

\begin{figure}
\includegraphics[width=84mm]{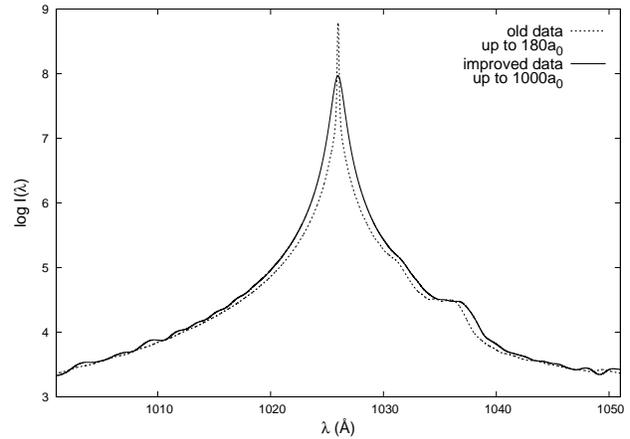}
\caption{Profile for the centre of the Lyman~$\beta$ line for a temperature of $10\,000$~K and perturber density of 10$^{17}$~cm$^{-3}$ evaluated using the data published in \citet{santos2012} (\textit{dashed line}) and the profile calculated with the improved data (\textit{continuous line}). The broadening of the line is underestimated when one uses the old data, affecting also the position of the satellite.}
\label{ly3cent}
\end{figure}

\begin{figure}
\includegraphics[width=84mm]{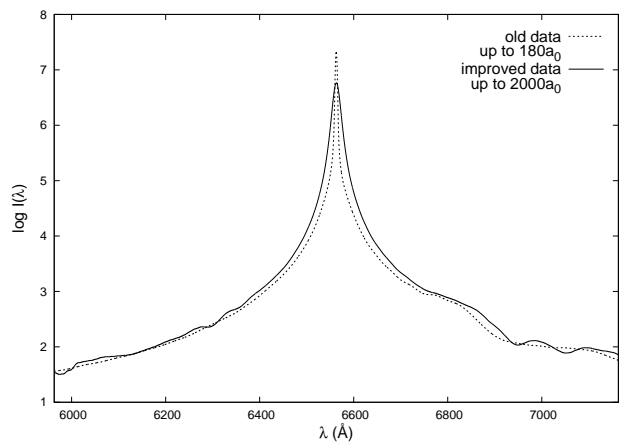}
\caption{Profile for the centre of the Balmer~$\alpha$ line for a temperature of $10\,000$~K and perturber density of 10$^{17}$~cm$^{-3}$ evaluated using the data published in \citet{santos2012} (\textit{dashed line}) and the profile calculated with the improved data (\textit{continuous line}). Again it is clear that the broadening is underestimated when the potential is truncated.}
\label{ba3cent}
\end{figure}

The obtained profile for Lyman~$\alpha$ is shown in both Fig. \ref{center1} and Fig. \ref{center2} to comparison with the profiles with previous data. It is extremely different than the one calculated with the old data without truncation, since it has no discontinuities (Fig. \ref{center1}). Compared to the profile calculated with the old data truncated in 180~a$_0$ (Fig. \ref{center2}), the difference is little, but it still important, especially in the central wavelengths, were we can see that the broadening is underestimated if one does not consider the behaviour at large distances. The same result is obtained when one analyses the profiles for Lyman~$\beta$ (Fig. \ref{ly3cent}) and Balmer~$\alpha$ (Fig. \ref{ba3cent}), with the addiction that it becomes apparent that such underestimation may also affect the position of the satellite, which will appear closer to the line centre.

\section{Conclusions}

The quasi-molecular contributions of H$_2^+$ are an effect very important to the shape of the lines in a hydrogen gas. The most remarkable contributions are the satellite lines, which are a consequence of the formation of transient bond states between hydrogen atoms and free protons due to an extremum in the potential difference. However, that is not the only effect of the quasi-molecular absorptions: they are also responsible for large asymmetries in the line wings. The satellites are an effect often discussed in the literature \citep[e.g.][]{allard1992, allard2009, santos2012}, since they are quite easily observed in the ultraviolet. Meanwhile, the asymmetry seems to be ignored, as it may not be directly detected. However, the fact that the quasi-molecular profile red wing falls down very quickly, especially for transitions involving states with high $n$, is probably very important in the opacity determinations. Not taking it into account, and continue to use the symmetric Stark profiles for protons, is almost certainly affecting not only the determinations of temperature, but also of gravity, to which such high order lines are more sensible.

The only line considered at present in this sense is Lyman~$\alpha$, whose quasi-molecular contributions turned out to be very important to the opacity of cool white dwarfs ($T_{\textrm{eff}}<6\,000$~K), as shown firstly by \citet{kowalski2006}. It is worth noticing that, at such temperatures, satellite lines are not even observed, but the quasi-molecular effect is important anyhow. The first step towards generalization of those results was done by \citet{santos2012}, who calculated all potentials energies for $n$ up to 10 and dipole moments for the Lyman series up to Ly~$\delta$ and the Balmer series up to H~10. They also evaluated the quasi-static profiles for those lines, and already pointed out there that the quasi-molecular effects in the line wings shall probably be important for opacity, but remain unaccounted for in the current models, which assume the interaction between hydrogen and protons is completely analogous to that of hydrogen with electrons. Considering that the quasi-static profiles are not correct at the line centres, we re-evaluated here all those line profiles, now using the unified theory, that can give the whole line shape. The data files are available to the community as online material accompanying this paper and also at our web page (http://astro.if.ufrgs.br/ingrid/index.htm). We believe that such profiles are enough to study both Lyman and Balmer series in white dwarfs.

We also noted that the precision of the current potential energy data was not enough to compare the calculated profiles to experimental results. The main reason for that were discontinuities due to numerical imprecision. We solve that in the code by increasing the truncation in our recurrence relations from index 25 to 300. Results were shown in section \ref{center} for Lyman~$\alpha$, Lyman~$\beta$ and Balmer~$\alpha$ and a large improvement could be verified. We hope to be able to compare our profiles to experimental data and actually confirm the importance of quasi-molecular contributions due to H$_2^+$ even when no satellites are observed, in order to make their inclusion in atmosphere models more consistent and consolidated.

\section*{Acknowledgments}

We thank Drs. D. Koester and R. D. Rohrmann for the enlightening discussions and useful suggestions throughout this work. I. Pelisoli and S. O. Kepler are supported by CNPq and FAPERGS-Pronex-Brazil.

\bsp

\label{lastpage}


\begin{thebibliography}{99}

\bibitem[\protect\citeauthoryear{Anderson}{1952}]{anderson1952} Anderson P. W., 1952, Phys. Rev., 86, 809--809
\bibitem[\protect\citeauthoryear{Andrews et al.}{2012}]{andrews2012} Andrews J.~J., Ag{\"u}eros M.~A., Belczynski K., Dhital S., Kleinman S.~J., West A.~A., 2012, ApJ, 757, 170
\bibitem[\protect\citeauthoryear{Allard}{1978}]{allard1978} Allard N. F., 1978, J. Phys. B: At. Mol. Phys., 11, 1383--1392
\bibitem[\protect\citeauthoryear{Allard \& Koester}{1992}]{allard1992} Allard N. F. , Koester D., 1992, A\&A, 258, 464--468
\bibitem[\protect\citeauthoryear{Allard, Sahal-Brechot \& Biraud}{1974}]{allard1974} Allard N. F., Sahal-Brechot S., Biraud Y. G., 1974, J. Phys. B: At. Mol. Phys., 7, 2158--2172
\bibitem[\protect\citeauthoryear{Allard et al.}{1994}]{allard1994} Allard N. F. , Koester D., Feautrier N., Spielfiedel A., 1994, A\&AS, 108, 417--431
\bibitem[\protect\citeauthoryear{Allard, Kielkopf \& Feautrier}{1998}]{allard1998} Allard N. F., Kielkopf J., Feautrier N., 1998, A\&A, 330, 782, 790
\bibitem[\protect\citeauthoryear{Allard et al.}{1999}]{allard1999} Allard N. F., Royer A., Kielkopf J. F., Feautrier, N., 1999, Phys. Rev. A, 60, 1021--1033
\bibitem[\protect\citeauthoryear{Allard et al.}{2000}]{allard2000} Allard N. F., Kielkopf J., Drira I., Schmelcher P., 2000, The European Physical Journal D, 12, 263--268
\bibitem[\protect\citeauthoryear{Allard, Kielkopf \& Loeillet}{2004}]{allard2004} Allard N. F., Kielkopf J. F., Loeillet B., 2004, A\&A, 424, 347--354
\bibitem[\protect\citeauthoryear{Allard et al.}{2004}]{allard2004b} Allard N. F., Kielkopf J. F., H\'{e}brard G., Peek J. M., 2004, The European Physical Journal D, 29, 7--16
\bibitem[\protect\citeauthoryear{Allard, Noselidze \& Kruk}{2009}]{allard2009} Allard N. F., Noselidze I., Kruk J. W., 2009, A\&A, 506, 993--997
\bibitem[\protect\citeauthoryear{Bergeron, Saffer \& Liebert}{1992}]{bergeron1992} Bergeron P., Saffer R. A., Liebert J., 1992, ApJ, 394, 228
\bibitem[\protect\citeauthoryear{Doherty et al.}{2014}]{doherty2014} Doherty C. L., Gil-Pons P., Siess L., Lattanzio J. C., Lau H. H. B., 2014, preprint (arXiv:1410.5431)
\bibitem[\protect\citeauthoryear{Falcon et al.}{2010a}]{falcon2010} Falcon R. E., Rochau G. A., Bailey J. E., Ellis J. L., Montgomery M. H., Winget D. E., Gomez M. R., Leeper R. J., 2010a, 17$^{\textrm{\tiny{th}}}$ European White Dwarf Workshop --- AIP Conference Proceedings, 1273, 436--439
\bibitem[\protect\citeauthoryear{Falcon et al.}{2010b}]{falcon2010b} Falcon R. E.,  Winget D. E., Montgomery M. H., Williams K. A., 2010b, ApJ, 712, 585--595
\bibitem[\protect\citeauthoryear{Falcon et al.}{2012}]{falcon2012} Falcon R. E., Rochau G. A., Bailey  J. E., Ellis J. L., Carlson A. L., Gomez T. A., Montgomery M. H., Winget D. E., Chen  E. Y., Gomez M. R., Nash T. J., 2012, High Energy Density Physics, 9, 82--90
\bibitem[\protect\citeauthoryear{Falcon et al.}{2014}]{falcon2014} Falcon R. E., Rochau G. A., , Bailey J. E., Gomez T. A., Montgomery M. H., Winget D. E., Nagayama T., 2014, preprint (arXiv:1410.4215F)
\bibitem[\protect\citeauthoryear{Genest-Beaulieu \& Bergeron}{2014}]{beaulieu2014} Genest-Beaulieu C., Bergeron P., 2014, preprint (arXiv:1410.5255)
\bibitem[\protect\citeauthoryear{Gianninas, Bergeron \& Ruiz}{2011}]{gianninas2011} Gianninas A., Bergeron P., Ruiz M.~T., 2011, ApJ, 743, 138
\bibitem[\protect\citeauthoryear{Giammichele, Bergeron, \& Dufour}{2012}]{giammichele2012} Giammichele N., Bergeron P., Dufour P., 2012, ApJS, 199, 29
\bibitem[\protect\citeauthoryear{Greenstein}{1980}]{greenstein1980} Greenstein J. L., 1980, ApJ, 241, L89--L93
\bibitem[\protect\citeauthoryear{Holm et al.}{1985}]{holm1985} Holm A. V., Panek R. J., Schiffer III F. H., Bond H. E., Kemper E., Grauer, A. D., 1985, ApJ, 289, 774
\bibitem[\protect\citeauthoryear{Hubeny}{1988}]{hubeny1988} Hubeny I., 1988, Computer Physics Comm., 52, 103
\bibitem[\protect\citeauthoryear{Hubeny \& Lanz}{2011}]{hubeny2011} Hubeny I., Lanz T., 2011, Astrophysics Source Code Library
\bibitem[\protect\citeauthoryear{Kepler et al.}{2006}]{kepler2006} Kepler S. O., Castanheira B. G., Costa A. F. M., Koester D., 2006, MNRAS, 372, 1799--1803
\bibitem[\protect\citeauthoryear{Kepler et al.}{2015}]{kepler2015} Kepler S. O., Pelisoli I., Koester D., Ourique G., Kleinman S. J., Romero A. D., Nitta A., Eisenstein D. J., Costa J. E. S., K\"{u}lebi B., Jordan S., Dufour P., Giommi P., Rebassa-Mansergas A., 2015, MNRAS, 446, 4078--4087
\bibitem[\protect\citeauthoryear{Kielkopf \& Allard}{1995}]{kielkopf1995} Kielkopf J. K., Allard N. F., 1995, ApJ, 450, L75--L78
\bibitem[\protect\citeauthoryear{Kielkopf, Allard \& Decrette}{2002}]{kielkopf2002} Kielkopf J. K., Allard N. F., Decrette A., 2002, The European Physical Journal D, 18, 51--59
\bibitem[\protect\citeauthoryear{Kleinman et al.}{2013}]{kleinman2013} Kleinman S. J., Kepler S. O., Koester D., Pelisoli I., Pe\c{c}anha V., Nitta A., Costa J. E. S., Krzesinski J., Dufour P., Lachapelle F.-R., Bergeron P., Yip C.-W., Harris H. C., Eisenstein D. J., Althaus L., C\'{o}rsico A., 2013, ApJS, 204:5
\bibitem[\protect\citeauthoryear{Koester et al.}{1985}]{koester1985} Koester D., Weidemann V., Zeidler K. T. E. M., Vauclair G., 1985, A\&A, 142, L5--L8
\bibitem[\protect\citeauthoryear{Koester et al.}{1996}]{koester1996} Koester D., Finley D. S., Allard N. F., Kruk J. W., Kimble R. A., 1996, ApJ, 463, L93
\bibitem[\protect\citeauthoryear{Kowalski \& Saumon}{2006}]{kowalski2006} Kowalski P. M., Saumon, D., 2006, ApJ, 641, L137--L140
\bibitem[\protect\citeauthoryear{Iben, Ritossa, \& Garc{\'{\i}}a-Berro}{1997}]{iben1997} Iben I., Jr., Ritossa C., Garc{\'{\i}}a-Berro E., 1997, ApJ, 489, 772
\bibitem[\protect\citeauthoryear{Liebert, Bergeron \& Holberg}{2005}]{liebert2005} Liebert J., Bergeron P., Holberg J. B., 2005, ApJS, 156, 47
\bibitem[\protect\citeauthoryear{Lindholm}{1945}]{lindholm1945} Lindholm E., 1945, Ark. Fys. A 32, No. 17, 1
\bibitem[\protect\citeauthoryear{Moehler \& Bono}{2008}]{moehler2008} Moehler S. \& Bono G., 2008, arXiv:0806.4456
\bibitem[\protect\citeauthoryear{Nelan \& Wegner}{1985}]{nelan1985} Nelan E. P., Wegner G., 1982, ApJ, 289, L31--L33
\bibitem[\protect\citeauthoryear{Rohrman, Althaus \& Kepler}{2011}]{rohrman2011} Rohrmann R. D., Althaus L. G., Kepler S. O., 2011, MNRAS, 411, 781--791
\bibitem[\protect\citeauthoryear{Romero et al.}{2012}]{romero2012} Romero A. D., C\'{o}rsico A. H., Althaus L. G., Kepler S. O., Castanheira B. G., Miller Bertolami M. M., 2012, MNRAS, 420, 1462--1480
\bibitem[\protect\citeauthoryear{Santos \& Kepler}{2012}]{santos2012} Santos M. G., Kepler S. O., 2012, MNRAS, 423, 68--79
\bibitem[\protect\citeauthoryear{Smartt et al.}{2008}]{smartt2008} Smartt S. J., Crockett R. M., Eldridge J. J., Maund J. R., 2008, IAU Symposium, 250, 201
\bibitem[\protect\citeauthoryear{Tremblay \& Bergeron}{2009}]{tremblay2009} Tremblay P.-E., Bergeron P., 2009, ApJ, 696, 1755
\bibitem[\protect\citeauthoryear{Tremblay et al.}{2013}]{tremblay2013} Tremblay P.-E., Ludwig H.-G., Steffen M., Freytag B., 2013, A\&A, 559, A104
\bibitem[\protect\citeauthoryear{Tremblay et al.}{2014}]{tremblay2014} Tremblay P.-E., Kalirai J. S., Soderblom D. R., Cignoni M., Cummings J., 2014, ApJ (accepted for publication)
\bibitem[\protect\citeauthoryear{Vidal, Cooper \& Smith}{1971}]{vcs1971} Vidal C.~R., Cooper J., Smith E.~W., 1971, J. Quant. Spectrosc. Radiat. Transfer., 11, 263--281
\bibitem[\protect\citeauthoryear{Wegner}{1982}]{wegner1982} Wegner G., 1982, ApJ, 261, L87--L89
\bibitem[\protect\citeauthoryear{Winget et al.}{1987}]{winget1987} Winget D. E., Hansen C. J., Liebert James, van Horn H. M., Fontaine G., Nather R. E., Kepler S. O., Lamb D. Q., 1987, ApJ, 315, L77--L81
\bibitem[\protect\citeauthoryear{Wolff et al.}{2001}]{wolff2001} Wolff B., Kruk  J. W., Koester D., Allard N. F., Ferlet R., Vidal-Madjar A., 2001, A\&A, 373, 674--682
\bibitem[\protect\citeauthoryear{Zygelman \& Dalgarno}{1990}]{zygelman1990} Zygelman B., Dalgarno A., 1990, ApJ, 365, 239--240

\end{thebibliography}
\end{document}